\documentstyle[aps,preprint]{revtex}
\begin{document}
\draft
\title{Discretized Superstring in Three Dimensional Superspace }
\author{Parthasarathi Majumdar}
\address{The Institute of Mathematical Sciences, CIT Campus, Madras
600113,  India.}
\maketitle
\begin{abstract}
The partition function of the discretized superstring in a target
superspace of three (Euclidean) bosonic dimensions, is shown, for a fixed
triangulation of the random world sheet, to be derived from the
partition function of a discretized bosonic string with an external
field present in the action in the form of a specific constant matrix,
using  first order forms of the actions. This latter partition
function appears more amenable to an exact analytical treatment.
\end{abstract}
The theory of strings in spacetimes of dimension lower
than the critical dimension stands far from complete. Most attempts
are stymied by the so-called $c_m \leq 1$ barrier,\cite{kpz,ddk}
where $c_m$ is the
central charge of the matter two dimensional conformal field theory
(CFT) on the
(random) string world sheet. In the continuum
approach, the conformal mode of the world sheet metric plays the role
of an additional target space dimension,\cite{das} so that $D=c_m+1$ for the
bosonic string. It follows that, for the susceptibility of the bosonic
non-critical string to have real critical exponents, $D \leq 2$. This
pathology manifests even more clearly in the discretized version of
the theory (by dynamical triangulations of the random 2d surface)
\cite{kaz,amb1,dav}, in that the lattice string tension does not scale
to zero at the critical point \cite{dft,amb2}. This implies
that the discrete random surface does not admit a continuum limit for
embeddings in (bosonic) spacetimes of $D>2$, possibly degenerating
into an ensemble of branched polymers (see, e.g., ref.\cite{amb3}).

In this letter, we consider an extension of the Polyakov string with
target space supersymmetry, by embedding the
2d random surface in a {\it superspace} with three bosonic (Euclidean)
directions. If a continuum limit exists for this theory, the Liouville
field will supply the fourth dimension of what may be a realistic
target spacetime. Here
the fermionic coordinates are target space $(SO(3))$ spinors but world sheet
scalars which are anticommuting. The resemblance with the $b-c$ ghost
system of the bosonic string vis-a-vis world sheet properties
leads to the possibility that these coordinates contribute negatively to
$c_m$, although, strictly speaking that really depends on their dynamics.
If however this is the case, then the relation
$D=c_m+1$ of the bosonic string will no longer be true, and one can
indeed have $D > 2$ while maintaining $c_m \leq 1$.\cite{nonu}

A second
motivation for considering such an extension comes from numerical work
by Ambjorn and Varsted \cite{amb4} wherein some improvements over the
bosonic case have
been obtained in the estimates for the ratio of the average
circumference and radius of the discretized world sheet. However,
these are subject to the usual restrictions due to limitations on
computer memory etc.  So an analytical handle on the partition
function appears desirable. A complete solution to this problem is
still beyond our reach; the following is hopefully a step in that
direction.

The partition function for a two dimensional triangulated random surface
(assumed to be of
spherical topology for simplicity) with matter is given as a sum over
triangulations $T$ as,\cite{amb1,dav}
\begin{equation}
Z~=~\sum_{T} \exp \{-\Lambda |{T}| \} \rho(T) Z_m^T~~ \label{bpar}
\end{equation}
where $\Lambda$ is the `cosmological constant' $|{T}|$ is the number of
triangles for a given triangulation, which equals the area of the
random surface; $\rho(T)$ is a symmetry factor, while $Z_m^T$ is the
partition function for matter for a given triangulation. For the
discretized version of
the bosonic Polyakov string, which corresponds to an embedding of the random
surface in a space of 3 (Euclidean) dimensions, the matter partition
function is given by,
\begin{equation}
Z_m^T~=~\int \prod_{\mu=1}^3 \prod_{i=1}^N  dX_i^{\mu}
\exp \{-\sum_{<{ij}>;\mu} (X_i^{\mu}~-~X_j^{\mu})^2 \}~~.
\label{bmat} \end{equation}

If we introduce the link variables $P_{ij}^{\mu}$,
defined to be antisymmetric in $i$ and $j$, the Gaussian action in (2)
can be linearized with respect to the $X_i^{\mu}$'s, giving the first
order form\cite{amb4}
\begin{equation}
S[X,P]~=~\frac12 \sum_{<{ij}>; {\mu}} \{ (P_{ij}^{\mu})^2 ~+~2 i
(X_i^{\mu}~- X_j^{\mu}) P_{ij}^{\mu} \}~.~\label{bfa} \end{equation}
The corresponding matter partition function can be rewritten after
integrating over the $X$ variables as
\begin{equation}
Z_m^T~=~\int \prod_{\mu;<{ij}>} dP^{\mu}_{ij} \exp \{
\sum_{\mu;<{ij}>}  (P^{\mu}_{ij})^2 \} \prod_{\mu;i} \delta (\sum_j
P^{\mu}_{ij})~ ~. \label{bfp} \end{equation}

In the supersymmetric case, the partition function for the first order
form of the action is given by\cite{amb4,mik}
\begin{equation}
{\tilde Z}_m^T~=~\int \prod_{\mu;<{ij}>} dP^{\mu}_{ij} \prod_{\mu;i}
dX^{\mu}_i
\prod_i d\theta_i \exp { - {\tilde S}[X, P; \theta ] }~~, \label{sfp}
\end{equation}
where,
\begin{equation}
 {\tilde S} [X, P; \theta]~=~\sum_{\mu;<{ij}>} \left [ (P^{\mu}_{ij})^2
{}~+~ 2i P^{\mu}_{ij} (X^{\mu}_i - X^{\mu}_j + i {\bar \theta_i}
\sigma^{\mu}  \theta_j )
\right]~~.\label{sfa} \end{equation}
The integrations over both the $X$ and the $\theta$ variables can be done
exactly, yielding\cite{amb4}
\begin{equation}
{\tilde Z}_m^T~=~\int \prod_{\mu;<{ij}>} dP^{\mu}_{ij}
e^{-\sum_{\mu;<{ij}>} (P^{\mu}_{ij})^2 } Det
\left(\sum_{\mu}(\sigma^{\mu} P^{\mu})
\right ) \prod_{\mu;i} \delta(\sum_j P^{\mu}_{ij} )~~, \label{sef}
\end{equation}
where,  the $P^{\mu}_{ij}$ for different $\mu$ are
assumed to commute, as in ref.\cite{amb4}. We now use a standard
formula for expressing a
$2N \times 2N$ determinant (viz. that obtained from the fermionic
integration ) in terms of an $N \times N$ determinant, to get
\begin{equation}
Det (\sum_{\mu} \sigma^{\mu} P^{\mu}_{ij}) ~=~Det (- \sum_{\mu} (P^{\mu})^2
)~~, \label{fdet} \end{equation}
so that,
\begin{equation}
{\tilde Z}_m^T~=~\int \prod_{\mu;<{ij}>} dP^{\mu}_{ij}
e^{-\sum_{\mu;<{ij}>} (P^{\mu}_{ij})^2 } Det
\left(-\sum_{\mu}(P^{\mu})^2
\right ) \prod_{\mu;i} \delta(\sum_j P^{\mu}_{ij} )~. \label{sep}
\end{equation}
It is tempting to interpret eq. (\ref{sep})
as the
partition function for a three-matrix model with a `non-singlet' delta
function constraint and a potential which is a sum of Gaussian and
logarithmic parts; recall however that
the $P_{ij}^{\mu}$ are link variables on the triangular lattice. Thus,
the interpretation as a random matrix model will go through provided
appropriate constraints are inserted into the integral. To this end,
we introduce the {\it adjacency} or {\it incidence} matrix,
$$ D_{ij}~\equiv~\left\{ \begin{array}{lll}
                -1 & \mbox{ if $i$ and $j$ are nearest neighbours} \\
                 0 & \mbox{otherwise}
             \end{array}
\right \} $$
With this definition the partition function (\ref{sep}) can be written as
\begin{equation}
{\tilde Z}_m^T~=~\int \prod_{\mu}{\cal D} P^{\mu} e^{Tr
\left \{ \sum_{\mu} (P^{\mu})^2 ~+~ ln [ \sum_{\mu} (P^{\mu})^2 ] \right \}
} \prod_{\mu;i\neq j} \delta \left ( (1-D_{ij}) P_{ij}^{\mu} \right )
\prod_{\mu; i} \delta( \sum_j P^{\mu}_{ij}) ~~. \label{mat} \end{equation}
Obviously, this is the partition function of a 3-matrix model with
`non-singlet' delta-function constraints. For
general random antisymmetric matrices, the same orthogonal
transformation, of course, does not diagonalize all three matrices
simultaneously. Notice that in this case the matrix integration space is
restricted to the space of independent fluctuating link variables,
rather than being totally random, due to imposition of the delta-function
constraint. The mutually commuting property of the $P^{\mu}$ for
different values of $\mu$ is valid on this constraint surface.

Consider now, the partition function
\begin{equation}
{\cal Z}(a_1,a_2, \dots, a_N)~\equiv~\prod_{\mu} \int {\cal D} P^{\mu}
e^{Tr [A (P^{\mu})^2] } \prod_{ij} \delta((1-D_{ij})P^{\mu}_{ij}) \prod_i
\delta(\sum_j P_{ij}^{\mu}) ~~, \label{maa} \end{equation}
where, the matrix $A$ is a real symmetric
$N \times N$ matrix defined as follows: for every orthogonal
transformation which
`diagonalizes' $P^{\mu}$ (i.e., diagonalizes the real symmetric matrix
$(P^{\mu})^2$ ) for any $\mu$, $A$ is defined by a
similarity transformation with the same orthogonal matrix on the
diagonal matrix $diag (a_1~a_2~\dots~a_N)~,~a_i~>~0$. In other words,
the matrix $A$ is classical in so far as its eigenvalues are
concerned, while its `angular' parts share the randomness of the
$P^{\mu}$.

With such a definition, the eq. (\ref{maa}) can be rewritten with some
abuse of notation, as
\begin{equation}
{\cal Z}(a_1,a_2, \dots, a_N)~=~\int \prod_{\mu} {\cal D} P^{\mu}
  e^{\sum_i [a_i \sum_{\mu}(p^{\mu}_i)^2] } \prod_{\mu;i,j}
\delta((1-D_{ij})P^{\mu}_{ij})  \prod_{\mu;i}
\delta(\sum_j P_{ij}^{\mu})~~, \label{mas} \end{equation}
where, $p_i^{\mu}$ are the eigenvalues of $P^{\mu}$. We have not
written out the integration measure in terms of integrations over the
eigenvalues and the angular variables of the orthogonal
transformation that diagonalizes $P^{\mu}$ and also the appropriate
Jacobian (the Van der Monde determinant), because we do not intend
to perform any explicit matrix integration here. The essential point
is that the exponential in the integrand is independent of the angular
variables.

Notice first that, the bosonic partition function
\begin{equation}
Z_m^T~~=~~{\cal Z}(1,1, \dots, 1)~~. \label{brec} \end{equation}
The point of importance now is that the supersymmetric partition
function (\ref{mat}) {\it can be derived} from (\ref{mas}), as we proceed
to show.

Consider
\begin{eqnarray}
\prod_i {\partial{} \over \partial{a_i}} {\cal Z}~ & = &~  \int
\prod_{\mu}  {\cal D} P^{\mu}
{\partial{} \over \partial{a_1}} {\partial{} \over \partial{a_2}}
\cdots  {\partial{} \over \partial{a_N}} e^{\sum_{\mu;i}
a_i (p^{\mu}_i)^2 }~\times~ \{constraints \} \nonumber \\
& = &~\int \prod_{\mu} {\cal D} P^{\mu} e^{Tr [A \sum_{\mu}(P^{\mu})^2] }
\prod_i [\sum_{\mu} (p^{\mu}_i)^2]~\times~ \{ constraints \} \label{spf}
\end{eqnarray}
where,
\begin{equation}
 \{ constraints \}~ \equiv ~ \prod_{\mu;i,j}
\delta((1-D_{ij})P^{\mu}_{ij}) \prod_{\mu;i} \delta(\sum_j P_{ij}^{\mu})~~
.\label{cns} \end{equation}
Since
\begin{equation}
\prod_i [\sum_{\mu} (p^{\mu}_i)^2 ]~=~Det [\sum_{\mu} (P^{\mu})^2]~~,
\end{equation}
which occurs in the supersymmetric partition function (\ref{mat}), we
observe that
\begin{equation}
{\tilde Z}_m^T~~=~~\prod_{i=1}^N {\partial{} \over \partial{a_i}}
{\cal Z}(a_1, \dots, a_N) |_{a_i=a_2=\cdots=a_N=1}~~. \label{man}
\end{equation}
This is the main result of the paper. Thus, if the partition function
(\ref{maa}) can be calculated exactly as a function of the
$\{ a_i \}$, then the supersymmetric partition function can indeed be
obtained analytically. The computation of ${\cal Z}(A)$ is in progress
\cite{pm} and will be reported elsewhere.

Notice that if ${\cal Z}(a_1,a_2, \dots a_N)$ is Taylor-expanded
around the point $a_i=1,~i=1,2, \dots N$ in powers of the $a_i$, then
the bosonic partition function is the first term in the expansion,
while the supersymmetric partition function corresponds to the
coefficient of the term $\prod_i a_i$ upto a factor of $N!$. It then
follows that this `parent' model represents a sort of generating
function for
discretized string theories and, as such, warrants investigation on
its own right. Similar one-matrix models for Hermitian matrices, with
an arbitrary non-singular
external matrix field were analyzed in the large $N$ limit in the
saddle point of approximation by Das et.
al. \cite{ddw} and also by Alvarez-Gaum\'e and Barb\'on \cite{alv}, as
attempts to construct higher dimensional matrix models of non-critical
strings. Our analysis has been restricted to the matter partition
function for a fixed triangulation and a very specific matrix $A$,
which might allow an exact solution for all $N$, notwithstanding the
complications of handling non-singlet delta function constraints.

Of course, the sum over triangulations will still remain to be
performed. The eventual aim envisages a computation of the lattice string
tension a l\'a ref.s {\cite{dft,amb2}}, and study of its scaling
behaviour near the critical point. This alone will enable one to
decide as to whether a continuum string theory exists without the
pathologies of the bosonic string.

To conclude, if a continuum limit does indeed exist, the target space
supersymmetry
will not extend into the extra `time' direction that emerges as a
transmutation of the
Liouville field, i.e., we shall only have a world
with three dimensional {\it space} supersymmetry. It is very likely that space
supersymmetry is adequate to produce a non-critical string theory with
a tachyon-free spectrum,\cite{kut}
while spacetime itself no longer has the full four dimensional
supersymmetry. The crucial issue in that case will be the vanishing of
the cosmological constant rather than a non-perturbative mechanism of
spacetime supersymmetry breaking.

\thanks{I am grateful to J. Ambjorn for bringing ref.{\cite{amb4}} to my
attention and also for a useful discussion. I thank R. Anishetty, R.
Basu, G. Date, S.
Govindarajan, N. D. Haridass, T. Jayaraman and V. John for useful
discussions. }


\begin{references}
\bibitem {nonu} { The corresponding
matter CFT may be non-unitary, but with coupling to
the random surface, the issue of unitarity is
nontrivial even for CFTs which are unitary in flat space. We donot
discuss this issue in this letter.}
\end{references}
\end{document}